# Point Island Models for Nucleation and Growth of Supported Nanoclusters during Surface Deposition

Yong Han[1], Émilie Gaudry[2], Tiago J. Oliveira[1,3], and James W. Evans[1,4,5]

[1]Department of Physics & Astronomy, Iowa State University, Ames, Iowa 50011
[2]Materials Department, Université de Lorraine, Campus Artem CS 14 234, 54 042 Nancy Cedex France
[3]Departamento de Física, Universidade Federal de Viçosa, 36570-900, Viçosa, Minas Gerais, Brazil
[4]Department of Mathematics, Iowa State University, Ames, Iowa 50011
[5]Ames Laboratory – USDOE, Iowa State University, Ames, Iowa 50011

**ABSTRACT**

Point island models (PIMs) are presented for the formation of supported nanoclusters (or islands) during deposition on flat crystalline substrates at lower submonolayer coverages. These models treat islands as occupying a single adsorption site, although carrying a label to track their size (i.e., they suppress island structure). However, they are particularly effective in describing the island size and spatial distributions. In fact, these PIMs provide fundamental insight into the key features for homogeneous nucleation and growth processes on surfaces. PIMs are also versatile being readily adapted to treat both diffusion-limited and attachment-limited growth, and also a variety of other nucleation processes with modified mechanisms. Their behavior is readily and precisely assessed by kinetic Monte Carlo simulation.

## 1. INTRODUCTION

Analysis of nucleation and growth in gas and condensed phases is a classic problem in chemical physics [1-3], the kinetics of which is traditionally analyzed in terms of Becker-Döring type rate equations. Studies continue both exploring the fundamental theory [4] and also including refinements to incorporate a more accurate molecular-level description of the process [5]. Already by the 1950's, the above treatments were refined to describe nucleation and growth during deposition on surfaces, and to account for the feature that often one finds small critical sizes for stable nuclei in these systems [6,7]. Such mean-field (MF) level treatments for surface deposition were developed and applied extensively over the next few decades [8]. However, only much later were certain fundamental shortcomings recognized [9]. These shortcomings derive from subtle spatial aspects of the nucleation and growth process on surfaces which are not incorporated, e.g., in simple Avrami type models [10].

We consider here nucleation and growth of nanoclusters (NCs) or islands during deposition on flat crystalline surfaces involving continuous random adsorption of atoms at a periodic array of adsorption sites together with rapid terrace diffusion of deposited atoms by hopping between neighboring adsorption sites [8,9]. The latter facilitates nucleation of new stable NCs in competition with growth of existing NCs. Characterization of these far-from-equilibrium systems involves two basic classes of questions. The <u>first</u> relates to the spatial and size distribution of NCs, i.e., NC ensemble-level properties. The second relates to the structure of individual NCs which depends sensitively on the details of the relaxation or rearrangement within NCs of aggregated atoms. Here, we focus on the first class of questions where the detailed structure of the individual NCs is not a significant factor.

To this end, we will consider tailored deposition models which simplify the treatment of NC structure by regarding these as "point islands" occupying a single adsorption site, but carrying a label to track their size (in atoms) [9,11]. See Fig.1. These so-called "point island models (PIMs)" are particularly valuable for two reasons. First, they are effective in elucidating fundamental aspects of the nucleation process, e.g., removing the complicating effect of the expanding island footprint. We emphasize that the simplicity of PIMs does not detract from their capability to capture basic features of the NC distribution. Second, this type of modeling is particularly versatile as it can be readily modified to incorporate a diverse range features and alternative pathways impacting nucleation. Kinetic Monte Carlo (KMC) simulation of these stochastic non-equilibrium lattice-gas models is also readily implemented with PIMs.

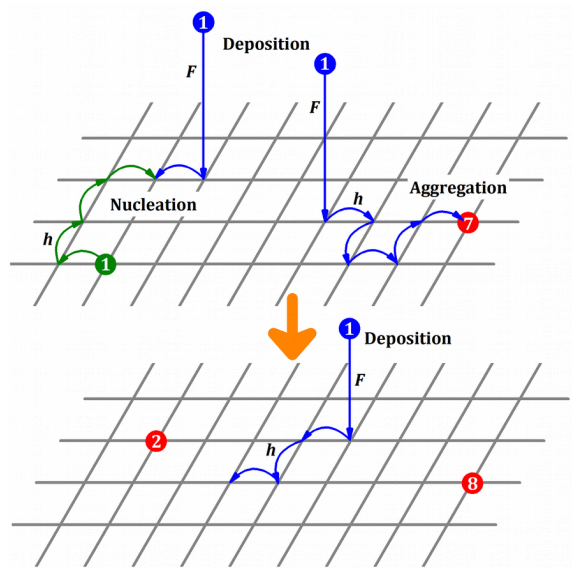

**Fig.1**. Schematic of the basic point island model (PIM) for critical island size $i = 1$. Circles indicate depositing atoms (blue), diffusing adatoms (green), and immobile islands (red), whose sizes are indicated by their numeric labels.

In Sec.2, we describe PIM for basic diffusion-limited homogeneous nucleation and growth processes, and review the analytic theory for such processes. We describe key features of the transition from transient to steady-state behavior which are important for our subsequent analysis. We also emphasize open challenges for a precise beyond-MF theory of NC size distribution. In Sec.3, we present a detailed analysis and new insights into scaling behavior for high surface diffusivity, as well as finite-size effects. Attachment-limited nucleation and growth is analyzed in Sec.4, both for a standard PIM and also for refinements mimicking formation of two-dimensional (2D) and three-dimensional (3D) islands. Various other refinements of the PIM to treat modified nucleation and growth processes are briefly described in Sec.5. Discussion of applications of various versions of the PIM to a variety of specific systems, as well as conclusions, are presented in Sec.6.

## 2. DIFFUSION-LIMITED HOMOGENOUS NUCLEATION & GROWTH

First, we specify the details of the PIM for diffusion-limited homogeneous nucleation and growth of NCs (also described as islands) during deposition [9,11]. Below T will denote the surface temperature, and we set $\beta = 1/(k_B T)$ where $k_B$ is the Boltzmann's constant. The surface is described by periodic square lattice of adsorption sites (although basic behavior is preserved for other lattices). See again Fig.1. Deposition occurs randomly at rate F per site. Terrace diffusion of deposited adatoms involves hopping to nearest-neighbor (NN) sites at rate h per direction. We will write $h = \nu \exp(-\beta E_d)$, where $E_d$ denotes the terrace diffusion barrier. As indicated above, islands occupy a single adsorption site, but carry a label to track their size, s, in units of atoms. With regard to nucleation and aggregation, there are two reasonable choices. We can specify that these processes are implemented when a hopping adatom reaches either: (A) the actual site on which another adatom or an island resides; or (B) a NN site to another atom or island. We should also note that nucleation and aggregation can also occur via direct deposition: (A) on the same site as the adatom or island; or (B) on a NN site. However, contributions from direct-deposition pathways are small, and thus are neglected in the analytic theory below which is focused on low island densities. Basic behavior will not depend strongly on the version (A) versus (B), and the difference between the models will vanish in the scaling limit $h/F \to \infty$ for fixed coverage $\theta = Ft$, where t denotes the deposition time. Results presented in Sec.3 correspond to version (B) and in Sec.4 to version (A). In the simplest implementation of this model, there is no diffusion of stable islands, no extra barrier for adatom attachment to islands, etc. These and other modifications are discussed in Sec.4 and 5.

Another key feature of the model is the specification of a critical size, i, such that only islands of $s > i$ atoms are stable, meaning that adatoms cannot detach from such islands [7-9]. Let $E_s < 0$ denote the binding energy for adatoms within a <u>sub-stable</u> island of size $s \leq i$, where $E_1 = 0$. One naturally chooses value for $E_s$ so that $|E_{s+1}| > |E_s|$ since larger clusters should have stronger overall binding than smaller clusters. Then, in the absence of attachment barriers (so h gives the attachment rate per direction), the rate of detachment per direction from an island of size s is given by $h \exp[-\beta(E_{s-1} - E_s)]$ according to detailed-balance. Actually, critical size i is realized only if $|\beta E_i|$ is not too large, so that detachment from islands of size i is facile on the time scale of aggregation. Otherwise, one has a smaller effective critical size. Furthermore, it is then expected that the population of sub-stable islands is quasi-equilibrated relative to the current density of diffusing adatoms (which reflects both the deposition flux and rates of nucleation and aggregation processes) [7,8,9]. In this case, behavior is strongly dependent on the binding energy, $E_i$, for the critical cluster, and effectively independent of $E_s$ for smaller $s < i$. We will focus on behavior for $i = 1$ which corresponds to irreversible island formation where behavior at a specified coverage, $\theta$, depends only on the ratio h/F. As an aside, we note that benchmark studies for $i > 1$ have often set $E_s = 0$ for $s \leq i$, so that detachment from sub-stable islands occurs at rate h [9].

The quantities of primary relevance in this study are the densities (per adsorption site) of adatoms, $N_1$, and of stable islands given by $N_{isl} = \Sigma_{s>i} N_s$, where $N_s$ is the density of islands of s atoms. The coverage satisfies $\theta = \Sigma_{s\geq 1} sN_s$ in units of monolayers (ML).

Figure 2 illustrates the two distinct regimes occurring during nucleation and growth of NCs during deposition [9,12,13]. Both regimes will feature in our analytic treatment, and can be described as follows:

(i) In the initial <u>transient regime</u>, $N_1 \approx Ft$ builds up uniformly over most of the surface leading to nucleation of far-separated islands at near-random locations. Depletion zones (DZs) form around each island. $N_1$ is reduced within the DZ due to aggregation with that island which can be regarded as a sink for diffusion adatoms. DZ radii increase like $R_{DZ} \sim (h\delta t)^{1/2}$, where $\delta t$ is the time since nucleation [12]. Later, growing DZs collide to form boundaries of capture zones (CZs) which surround each island, as described below. These boundaries are constructed to be roughly equidistant from the relevant point islands (a non-trivial construction process, the details of which are not critical here).

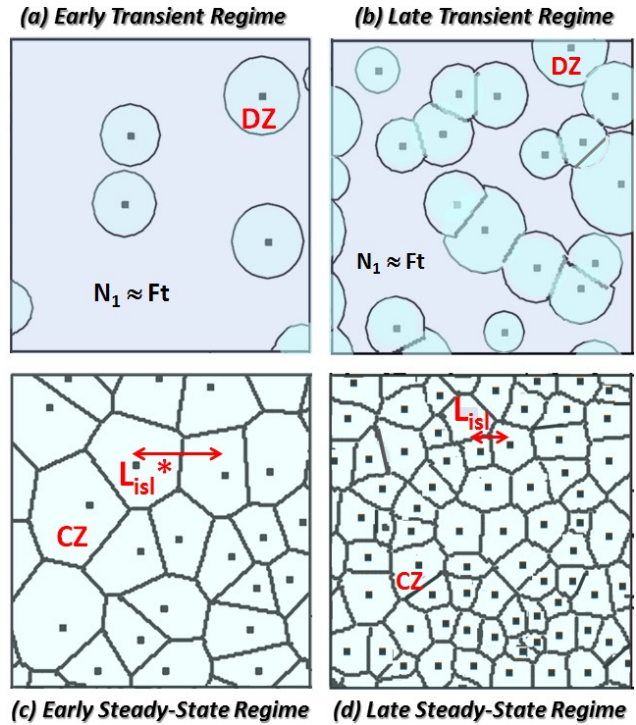

**Fig.2**. Schematic of key stages of nucleation and growth for irreversible island formation. (a, b) Transient regime with nucleation at quasi-random locations. Depletion zones (DZs) expand about just nucleated islands, and collide to form capture zones (CZs); (c, d) Steady-state regime in which most nucleation occurs. Behavior shown is for $i = 1$.

(ii) In the subsequent <u>steady-state regime</u>, DZs have expanded sufficiently so that the surface is completely covered or tessellated by the corresponding CZs surrounding each island. The basic feature of CZs is that most deposited atoms aggregate with the island associated with the CZ in which they are deposited, and ideally the growth rate of each island should be exactly proportional to the corresponding CZ area. Consequently,

CZs must strictly be constructed by solving a boundary value problem for deposition-diffusion equation, but in practice CZs for point islands are reasonably described by Voronoi cells based on the island distribution [9]. We emphasize that island nucleation persists and is dominant in the steady-state regime especially for small $i$.

In Sec.2A, we briefly review MF rate equation analysis for behavior of the average island density [8,9]. We follow in Sec.2B with an analysis of the island size distribution (ISD), where MF treatments fail qualitatively to predict the correct behavior, and where open challenges for reliable beyond-MF treatments remain [9].

## 2A. RATE EQUATIONS FOR THE AVERAGE ISLAND DENSITY

Traditional MF rate equations are effective for describing behavior of average island and adatom densities in 2D surface systems [8,9], despite ignoring logarithmic corrections discussed in Appendix A. Refinement is needed for 1D systems as discussed in Sec.5 and also Appendix A. We first describe the two key rates. The nucleation rate is given by $K_{nuc} = \sigma_i h N_1 N_i$, where the capture number, $\sigma_i$, describes the propensity of critical clusters to capture of diffusing adatoms. The aggregation rate is given by $K_{agg} = \sigma_{av} h N_1 N_{isl}$, where $\sigma_{av}$ is the average capture number for stable islands. The mean adatom and island densities satisfy

$$dN_1/dt = F - (i+1)K_{nuc} - K_{agg}, \text{ and } dN_{isl}/dt = K_{nuc}. \tag{1}$$

These equations are closed by invoking a quasi-equilibrium Walton relation [7] for the density of critical clusters $N_i \approx \exp[-\beta E_i](N_1)^i$. Integration of (1) reveals the transient regime where $d/dt\, N_1 \approx F$ so $N_1 \approx Ft = \theta$, followed by crossover to a steady-state regime. The latter reflects a balance between a gain of adatoms due to deposition and loss primarily due to aggregation, so that $F \approx K_{agg}$, and $N_1 \approx F/(\sigma_{av} h N_{isl})$. Substituting this form for $N_1$ into (1) for $N_{isl}$ and integrating yields (3a) and (3c) below. Matching the transient and steady-state solutions for $N_{isl}$ reveals crossover at (low) coverage of

$$\theta^* \sim \exp[\beta E_i/(i+3)](h/F)^{-2/(i+3)}. \tag{2}$$

Specifically, one finds that

$$N_{isl} \approx (i+2)^{-1}\sigma_i\, \theta^{i+2} \exp[-\beta E_i](h/F) \tag{3a}$$

in the transient regime,

$$N_{isl}^* \sim \exp[-\beta E_i/(i+3)](h/F)^{-\chi^*} \tag{3b}$$

at crossover where $\theta \approx \theta^*$, and

$$N_{isl} \approx [(i+2)\sigma_i/(\sigma_{av})^{i+1}]^{1/(i+2)}\, \theta^{1/(i+2)} \exp[-\beta E_i/(i+2)](h/F)^{-\chi} \tag{3c}$$

for $\theta = O(1)$, the latter being the steady-state regime. The scaling exponents satisfy

$$\chi = i/(i+2) \text{ versus } \chi^* = (i+1)/(i+3), \text{ so } \chi^* - \chi = 2(i+2)^{-1}(i+3)^{-1} > 0. \tag{4}$$

As an aside, one can immediately calculate the average island separation at crossover from $L_{isl}^* = (N_{isl}^*)^{-1/2}$ and in the steady-state from $L_{isl} = (N_{isl})^{-1/2}$. Also see Fig. 2 for $L_{isl}$ and $L_{isl}^*$.

These results immediately yield several fundamental insights into the behavior of $N_{isl}$ and of related quantities. <u>First</u>, consider the key result (3c) for steady-state $N_{isl}$. The F-dependence of $N_{isl}$ determines χ and thus i. The Arrhenius energy of $N_{isl}$ is given by $E = (iE_d - E_i)/(i + 2)$, so that $E = E_d/3$ for $i = 1$ determining $E_d$. <u>Second</u>, a particularly significant observation is that the steady-state $N_{isl}$ far exceeds its crossover value $N_{isl}*$ for large h/F since χ*-χ > 0, dramatically so for smaller i. Correspondingly, $L_{isl}*$ far exceeds $L_{isl}$. The predominance of nucleation in the steady-state regime for large h/F is quantified by the ratio

$$N_{isl}/N_{isl}* \sim (\theta/\theta^*)^{1/(i+2)}. \tag{5}$$

<u>Third</u>, we consider the behavior of $N_1$. One can use the steady-state relation $N_1 \approx F/(\sigma_{av}hN_{isl})$ together with (3c) to obtain

$$N_1 \sim \theta^{-1/(i+2)} \exp[+\beta E_i/(i+2)](h/F)^{-2/(i+2)} \tag{6}$$

in the steady-state. By matching (6) to $N_1 \approx \theta$ in the transient regime, one immediately recovers the result (2) for the crossover θ*. From (6), it is also clear that $N_1 << N_{isl}$ for $i = 1$, $N_1 \sim N_{isl}$ for $i = 2$, and $N_1 >> N_{isl}$ for $i > 2$ when $\theta = O(1)$ for large h/F [14]. <u>Fourth</u>, we consider the mean island size, $s_{av}$. If most adatoms (or even a finite fraction of them) are incorporated into islands, then one has $s_{av} \sim \theta/N_{isl}$, so that

$$s_{av} \sim \theta^{(i+1)/(i+2)} \exp[+\beta E_i/(i+2)](h/F)^{i/(i+2)} \tag{7}$$

in the steady-state. It is common to write $s_{av} \sim \theta^z$ and $N_{isl} \sim \theta^{1-z}$ with dynamic scaling exponent $z = (i+1)/(i+2)$. One can show that a finite fraction of adatoms are incorporated into islands at crossover, so that $s_{av}* \sim \theta/N_{isl}*$ applies there also. It then follows that the corresponding $s_{av}*$ is far below $s_{av}$ for smaller i and large h/F, with $s_{av}* = O(1)$ for $i = 1$, so that island growth up to crossover is "negligible". <u>Fifth</u>, one can show that the nucleation rate has the form

$$K_{nuc}/F \sim (h/F)^{-(i-1)/(i+3)} \exp[-2\beta E_i/(i+3)] k(\theta/\theta^*), \tag{8}$$

where $k(u) \sim u^{i+1}$ [15] in the transient regime, and $k(u) \sim u^{-(i+1)/(i+2)}$ in the steady-state regime. Thus, k(u) and the nucleation rate are strongly peaked around crossover $u \approx 1$. The universal nature of the prefactor in front of k(θ/θ*) for both transient and steady-state regimes follows since the distinct forms of $N_1$ for these regimes match at $\theta \approx \theta^*$.

## 2B. BEYOND-MF RATE EQUATIONS FOR THE ISLAND SIZE DISTRIBUTION

Of primary importance in analysis of the ISD is the rate at which diffusing adatoms aggregate with an island of size s, denoted by $K_{agg}(s) = \sigma_s h N_s N_1$ where $\sigma_s$ is the "capture number" for islands of size s. One then obtains [8,9]

$$dN_s/dt \approx K_{agg}(s-1) - K_{agg}(s), \tag{9}$$

for $s > i$. To recover the simpler reduced equation (1) for $N_{isl}$, one sums (9) over $s > i$ to obtain $dN_{isl}/dt \approx K_{agg}(i) = K_{nuc}$. We do not give details here, but one can also show that

$$K_{agg} = \Sigma_{s>i} K_{agg}(s) = \sigma_{av} h N_1 N_{isl}, \tag{10}$$

where $\sigma_{av} = \Sigma_{s>i}\sigma_s N_s/\Sigma_{s>i} N_s$.

We will focus on analysis of <u>scaling solutions</u> to (9) for large $h/F$ or large $s_{av}$ in the steady-state regime. Introducing a scaled island size variable, $x = s/s_{av}$, we anticipate that these solutions have the form [9,11,16]

$$N_s \approx (N_{isl}/s_{av})\, f(x), \tag{11}$$

for large $s_{av}$. Here, one has that $\int_{x>0} dx\, f(x) = \int_{x>0} dx\, xf(x) = 1$. The PIM is special in that $f(x)$ has no explicit $\theta$-dependence, a feature which is lost for islands of finite extent. Our analysis also introduces a scaling function for the capture numbers, $\sigma_s/\sigma_{av} \approx c(x)$ [17]. Then, we perform a quasi-hydrodynamic coarse-graining analysis of (9) treating rescaled island size as a continuous variable. Also exploiting the temporal scaling, $s_{av} \sim (Ft)^z$, and the steady-state relation for $N_1$, one obtains [16]

$$f(x) = f(0)\, \exp\{ \int_{0<y<x} dy\, [(2z-1) - dc(y)/dy]/[c(y) - zy] \}, \tag{12}$$

with the auxiliary relation $c(x=0)f(x=0) = 1 - z > 0$. See also Ref.[17]. The auxiliary relation shows that $f(x=0) > 0$ in qualitative contrast to the most commonly assumed heuristic Amar-Family form for the scaling function $f_{AF}(x) \propto x^i \exp\{-ia_i x^{1/a_i}\}$, where $\Gamma[(i+2)a_i]/\Gamma[(i+1)a_i] = (ia_i)^{a_i}$. [18]

One advantage of the PIM is that in a traditional MF theory, islands of any size, $s$, have the same capture number, as they have the same spatial extent. This in turn implies that $c_{MF}(x) \equiv 1$. In this case, (12) exhibits a singularity in the integral yielding the explicit non-analytic form [9,11]

$$f_{MF}(x) = (i+2)^{-1} H(x_i - x)\, (1 - x/x_i)^{-i/(i+1)} \text{ where } x_i \equiv (i+2)/(i+1), \tag{13}$$

where $H(u) = 0$ (1) for $u<0$ ($u>0$) is the Heaviside step function. In fact, (13) is also qualitatively incorrect since the exact $c(x)$ increases in a way which avoids this type of MF singularity.

It is clear that the above formulation (12) for the ISD provides an exact but incomplete theory since the form of $c(x)$ is not yet specified. Certainly, adopting a simple MF form $c_{MF}(x) \equiv 1$ is inadequate. Perhaps the greatest open challenge is to provide a quantitative beyond-MF theory for $c(x)$. We emphasize that any such formulation must account in detail for the subtle spatial aspects of nucleation during the steady-state regime. These spatial aspects, together with the persistent nature of nucleation in the steady-state regime, control the form of $c(x)$, and thus the form of $f(x)$.

Here, we just briefly indicate the current status of this effort. To this end, it is useful to adopt a geometric picture of adatom capture by islands, as already implied in the CZ picture of Fig.1. The idea is that the CZ area measures the rate of aggregation with an island, so that $K_{agg}(s)/N_s = \sigma_s hN_1 = FA_s$, where $A_s$ denotes the mean area of CZs for islands of size $s$. [8,9,16] It is also expected that CZs for PIM constructed as Voronio cells will reasonably describe the true CZs for which CZ area exactly describes island growth rate [9,19]. Then, introducing a scaling function for $A_s$ via $A_s = A_{av}\, a(x)$, where $A_{av} = 1/N_{isl}$ is the mean CZ area, it follows that $a(x) \equiv c(x)$. Thus, the most effective theories for beyond-MF $c(x)$ are actually formulated as rate-equation-type theories for $A_s$, and thus for $a(x)$. Indeed, one can write down a heuristic rate equation for the fractional CZ area, $A_sN_s$, associated with islands of size s accounting readily for gain and loss due to island growth, and less easily for loss associated with nucleation of new islands whose CZs "cut into" this area [20]. The most sophisticated approach starts with the joint probability distribution, $N_{s,A}$, for islands of size s and CZ area A [21-25]. A moment analysis leads to a refinement of the above-mentioned heuristic rate equation for $A_s$ which in turn leads to an equation for the scaling function $a(x)$. However, this equation requires as input, information on subtle spatial aspects of nucleation such as the probability that the CZ of a new island overlaps that of an existing island of size $s$, and also the typical amount of overlap [24].

## 3. SCALING AND FINITE-SIZE EFFECTS FOR THE DIFFUSION-LIMITED REGIME

### 3A. SCALING WITH h/F AND θ

The predictions of the MF rate equation theory of Sec.2A for the scaling behavior of $N_{isl}$ with $h/F$ and with θ have been confirmed by precise KMC simulation analysis at least for $i = 1$ and $i = 2$ [9,11]. We note, however, that there are corrections to ideal asymptotic scaling due to finite $h/F$, and due to logarithmic corrections [9,11]. (Logarithmic factors were not included in the capture numbers in our analysis, although such refinement is readily implemented. See Appendix A.) Thus, here we focus on more delicate scaling issues for the ISD and related quantities.

Deviations from asymptotic large-$h/F$ scaling for the ISD are more severe than for $N_{isl}$, not surprisingly since this quantity is more sensitive to the details of the nucleation and growth process and is not described by a simple MF theory. Previous studies have tended to explore approach to the scaling limit for fixed rather low θ ~ 0.1 ML and increasing $h/F$. These studies suggested that the ISD achieves the limiting scaling form somewhat slowly only when $h/F$ is above $10^9$ (corresponding to rather demanding simulations for large system sizes). This feature is illustrated in Fig.3a. For an assessment of convergence to the scaling limit of various quantities describing the shape $f(x)$ of the scaled ISD, in Fig.4 we show the coefficient of variation, $C = (<s^2>_c)^{1/2}/<s>$, the skewness, $S = <s^3>_c/(<s^2>_c)^{3/2}$, and the (excess) kurtosis, $K = <s^4>_c/(<s^2>_c)^2$. Here, $<s^n>_c$ denotes the $n^{th}$ cumulant of the ISD, where $<s^2>_c = <(s-<s>)^2>$, $<s^3>_c = <(s-<s>)^3>$, $<s^4>_c = <(s-<s>)^4> - 3(<s^2>_c)^2$. $C$ gives the standard deviation of $f(x)$ in the scaling limit, $S$ describes lack of reflection symmetry about $s = <s>$, and $K$ measures the weight of the distribution in the tails relative to a Gaussian distribution. Simple extrapolation to $h/F \to \infty$ indicates that $C_\infty \approx 0.46$ to $0.47$, $S_\infty \approx -0.32$ to $-0.41$, and $K_\infty \approx -0.83$ to $-0.89$. For comparison, for the MF scaling function $f_{MF}(x)$, one has $C_\infty = 1/\sqrt{5} \approx 0.447$, $S_\infty = -2\sqrt{5}/7 \approx -0.639$, and $K_\infty = -6/7 \approx -0.857$.

To provide a framework to understand convergence to the scaling limit (and also to potentially facilitate alternative analysis of this limit), we suggest that deviations from the asymptotic form of the ISD are controlled by the extent of persistent nucleation in the steady-state regime (beyond the transient regime). This extent is naturally quantified by the ratio $N_{isl}/N_{isl}{}^* \sim (θ/θ^*)^{1/(i+2)}$, and thus equivalently by the ratio $θ/θ^*$. Thus, since $θ^* \sim (h/F)^{-2/(i+3)}$, proximity to the scaling limit is reflected by the magnitude of the combination $\mathfrak{R} = θ^{(i+3)/2}(h/F)$. Asymptotic behavior is usually probed for smaller θ = 0.1 ML, say, by increasing h/F to at least $10^9$. The above observations suggest that asymptotic behavior can instead be probed retaining moderate $h/F \sim 10^6$, say, by sufficiently increasing θ. Support for this idea is provided by results in Fig.3b which show that the shape of the ISD for $i = 1$ as θ increases from 0.05 to 0.5 ML for fixed $h/F = 10^6$ appears to approach the asymptotic scaling form for $h/F \to \infty$. Specifically the shape for θ = 0.5 ML and $h/F = 10^6$ corresponds closely to that for θ = 0.1 ML and $h/F = 2.5 \times 10^7$ where for $i = 1$ one has $\Re = \theta^2(h/F) = 2.5 \times 10^5$ in both cases.

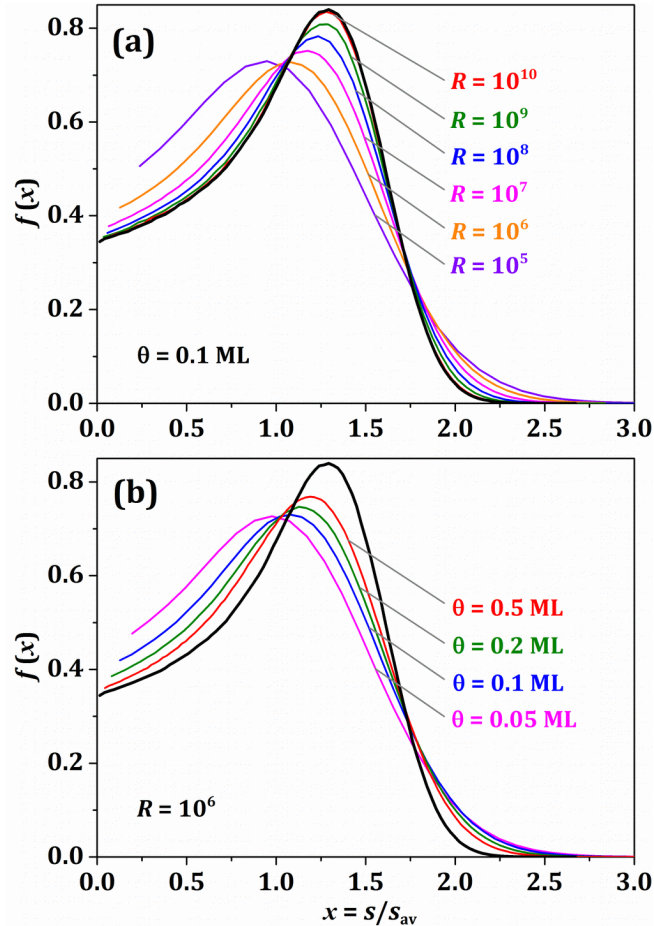

**Fig.3.** $f(x)$ for $i=1$. (a) θ=0.1 ML and $R = 4h/F$ increasing from $10^5$ to $10^{10}$; (b) $R = 10^6$ and θ increasing from 0.05 to 0.5 ML. Limiting distribution (thicker curve) is estimated from $R = 10^{10}$ and θ = 0.5 ML.

Finally, we mention that one can of course perform more detailed analysis of the approach to the scaling limit for the ISD and various cumulant ratios indicated in Fig.4 describing its shape. In addition to scaling with increasing $h/F$, one can consider scaling with increasing θ. However, a definitive characterization of behavior is difficult even with our current extensive data. Analysis for $K$ either increasing $h/F$ or θ yields reasonably consistent results, and a well-defined limiting value $K_\infty$. Analysis of S is more complex due to a slower approach to limiting behavior as $h/F \to \infty$. We leave further analysis and discussion of these subtleties for a separate publication.

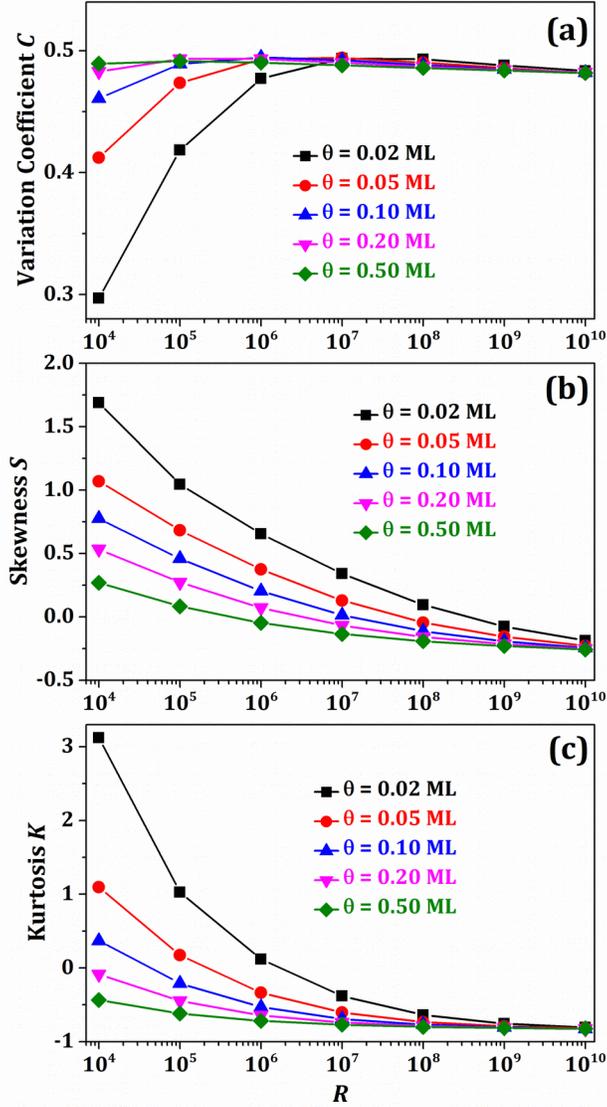

**Fig.4**. Variation with $R = 4h/F$ of the coefficient of variation, skewness, and (excess) kurtosis, for the ISDs for $i$ = 1. Behavior is shown for θ from 0.02 to 0.5 ML.

## 3B. FINITE-SIZE EFFECTS

In practice in performing simulations, one analyzes behavior for finite $L \times L$ site systems with periodic boundary conditions. One hopes that the side length, $L$ (in units of surface lattice constant), is large enough to avoid finite-size effects. In general, one expects such effects to become significant when the system size, $L$, becomes comparable to an intrinsic characteristic length. One potential complication for the considered systems here is that in addition to the natural characteristic length, $L_{isl}$, for θ

= O(1), there is a significantly larger length, $L_{isl}^*$. Certainly, choosing system size $L \ll L_{isl}^*$ will greatly impact system evolution in the transient and crossover regimes. The initial transient regime will no longer involve island nucleation at essentially random locations. However, we will show that for such sizes, behavior of at least basic quantities in the steady-state regime is not affected provided that $L \gg L_{isl}$. Results shown in Fig.5 for the island density in a finite system, $N_{isl}(L)$, relative to $N_{isl}(L = \infty)$ do reveal clear scaling with $L/L_{isl}$. This contrasts previous speculation [9,13]. Furthermore, limiting behavior is achieved quite quickly already by $L \approx 4L_{isl}$. As an aside, defining $L_{isl}$ based on either finite ($L_{isl} = [N_{isl}(L)]^{-1/2}$) or infinite ($L_{isl} = [N_{isl}(L = \infty)]^{-1/2}$) systems, or using the asymptotic form ($L_{isl} = \theta^{-1/6} (4h/F)^{-1/6}$), yields similar data collapses. See Fig. 5b. The last choice of $L_{isl}$ has practical utility, since from the results in Fig.5b, it allows ready assessment of the minimal $L$ which must be used in simulations, for given $\theta$ and $h/F$, to avoid finite-size effects. One might anticipate more complex behavior for more "delicate" quantities characterizing the ISD shape, but in Appendix B we show that these also exhibit simple scaling with $L/L_{isl}$.

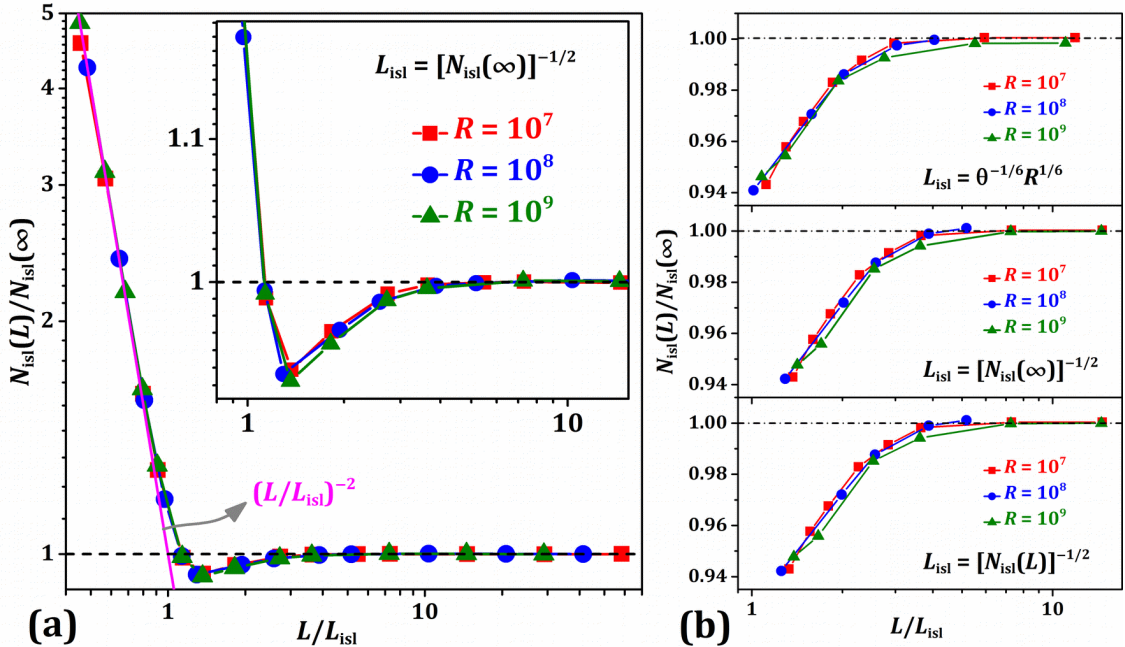

**Fig.5.** (a) Behavior of $N_{isl}(L)/N_{isl}(\infty)$ versus $L/L_{isl}$ for $L_{isl} = [N_{isl}(\infty)]^{-1/2}$ at $\theta = 0.1$ ML. The pink curve shows $N_{isl}(L)/N_{isl}(\infty) = (L/L_{isl})^{-2}$. The inset focuses on behavior near the minimum. (b) Comparison of scaling behavior for different choices of $L_{isl}$. Results are shown for $R = 4h/F = 10^7$ (red), $10^8$ (blue), and $10^9$ (green) at $\theta = 0.1$ ML.

To analyze behavior for $L/L_{isl} = O(1)$ in more detail, we introduce the probability $P_n(\theta)$ that there are <u>at least</u> n islands at coverage $\theta$ for a finite $L \times L$ site system. Equivalently, one can consider the probability, $p_n = P_n - P_{n+1}$, for <u>exactly</u> n islands (with $p_0 = 1 - P_1$). Then, one has that

$$N_{isl}(L)/N_{isl}(\infty) = (L/L_{isl})^{-2} <n>, \text{ where } <n> = \sum_n n p_n = \sum_n P_n. \qquad (14)$$

For small $L < L_{isl}$, one expects a "single island regime" with $P_1 \approx 1$ and $P_{n\geq 2} \approx 0$, so that $<n> \approx 1$ and $N_{isl}(L)/N_{isl}(\infty) \approx (L/L_{isl})^{-2}$, as confirmed in Fig. 5. Moreover, one anticipates that as $L$ increases above $L_{isl}$, the populations of $P_{n\geq 2}$ increase quickly. This behavior is supported by the simulation results for $P_n$ for various $L/L_{isl}$ shown in Fig.6a.

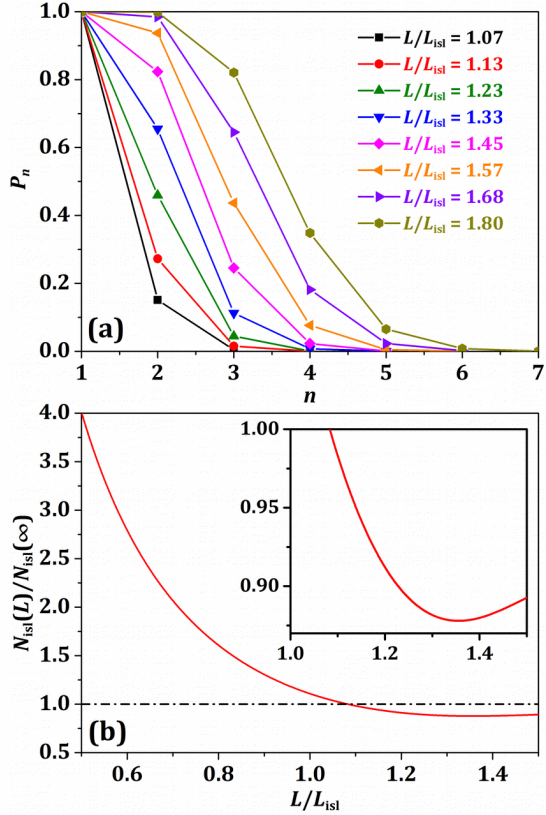

**Fig. 6**. (a) Probability $P_n$ for at least n islands nucleated for various $L/L_{isl}$ at $\theta=0.1$ ML and $R = 4h/F = 10^7$. (b) Analytic theory for crossover from the single-island regime. The curve is obtained using analytic expressions for $P_1 \approx 1$, $P_2$, and $P_3$ (from Appendix C), and setting $P_{n\geq 4} = 0$. The inset highlights the minimum in $N_{isl}(L)/N_{isl}(\infty)$ versus $L/L_{isl}$.

Further elucidation of the behavior shown in Fig.5 comes from making analytic estimates for $P_n$, at least for small $n$, and thus for $N_{isl}(L)$ for $L \sim L_{isl}$ and $\theta = O(1)$. To this end, we start by considering nucleation of the first island. The probability, $Q_1$, that no island exists at coverage $\theta$ (where $Q_1 = p_0$) is given by

$$Q_1(\theta) = \exp[-\int_{0<\theta'<\theta} d\theta'\ K_{Tnuc0}(\theta')/F], \qquad (15)$$

where $K_{Tnuc0}(\theta)/F = \sigma_i(h/F)\exp(-\beta E_i)\theta^{i+1}L^2$. Here, $K_{Tnuc0}(\theta)$ denotes the total nucleation rate in the $L \times L$ system (with periodic boundary conditions) at coverage $\theta$ in the absence of any island. This result shows that choosing $L$ to correspond to $L_{isl}$ for $\theta = O(1)$, the first island is nucleated at a coverage

$$\theta_{nuc}(1) \sim \exp[\beta(i+1)E_i/(i+2)^2](h/F)^{-2(i+1)/(i+2)^2}. \qquad (16)$$

For large $h/F$, $\theta_{nuc}(1)$ is just above $\theta^*$, and far below $\theta = O(1)$. Thus, for $L \sim L_{isl}$, the first island is nucleated very early, and $P_1 = 1 - Q_1 \approx 1$ for most $\theta$ above small $\theta_{nuc}(1)$.

After nucleation of the first stable island, a steady-state adatom density, $N_1(ss)$, is quickly established in the finite system. The basic features of $N_1(ss)$ follow from solving $F + h\nabla^2 N_1(ss) = 0$ in an annular domain with inner radius $r_{in} = 1$ and outer radius $r_{out} \approx L/2$ [26]. From this solution, one finds a typical magnitude for $N_1(ss) \sim (h/F)^{-1}L^2$ revealing that the total nucleation rate, $K_{Tnuc1}$, for this single-island system scales like

$$K_{Tnuc1}/F \sim \sigma_i \exp(-\beta E_i)(h/F)^{-i}L^{2(i+2)} \sim \sigma_i (L/L_{isl})^{2(i+2)}, \qquad (17)$$

where this quantity is independent of time or $\theta$. Then, the probability that the <u>second island</u> has not been nucleated at coverage $\theta$ is

$$Q_2(\theta) \approx \exp[-\theta K_{Tnuc1}/F] \approx \exp[-b\theta(L/L_{isl})^{2(i+2)}], \text{ and } P_2 = 1 - Q_2, \qquad (18)$$

with $b = O(1)$. The existence of the <u>third island</u> will depend on the time-independent total nucleation rate, $K_{Tnuc2}$, for the third island in the presence of two existing islands. Actually, $K_{Tnuc2}$ depends on the position of the second island relative to the first. However, it is clear that $K_{Tnuc2}$ scales in the same way as $K_{Tnuc1}$ with respect to $L/L_{isl}$. The same applies for nucleation of subsequent islands. See Appendix B.

For $L$ close to $L_{isl}$ with significant probability of a small number of islands, then one has that $<n> \approx P_1 + P_2 + \ldots$, and consequently that

$$N_{isl}(L)/N_{isl}(\infty) = g(L/L_{isl}), \text{ where } g(u) \approx u^{-2}\{1 + [1 - \exp(-b\theta u^{2(i+2)})] + \ldots\}. \qquad (19)$$

The corresponding behavior with $b = 1.1$ (consistent with behavior of $P_2$ in Fig.6a for $L/L_{isl}$ just above unity) is shown in Fig.6b also including the contribution from $P_3$.

Analysis of the crossover regime for higher $L/L_{isl} \approx 1.5 - 4$ is challenging. Nucleation rates becomes more complex as they depend on the locations of the multiple previously nucleated islands (which are not random as new islands prefer to be nucleated far from existing islands). However, from a more general perspective, just as convergence to the scaling limit in Sec.3A is controlled by $N_{isl}/N_{isl}^*$, one might expect that convergence to $L = \infty$ behavior reflects the ratio of the number of islands for $L > L_{isl}$ to that when $L = L_{isl}$ which is unity. This ratio is given by $L^2 N_{isl} = (L/L_{isl})^2$.

## 4. ATTACHMENT-LIMITED HOMOGENEOUS NUCLEATION & GROWTH

### 4A. PIM WITH EXTRA BARRIERS FOR NUCLEATION AND AGGREGATION

Our treatment of the effect of additional barriers on nucleation and growth [27,28] refines standard PIM. Here, for simplicity, we just describe these changes for the case of irreversible island formation $i = 1$, and below we only consider this case. We now include separate hopping rates $h_0 = h$ for terrace diffusion, $h_1 = h \exp(-\beta\delta^*)$ for nucleation, and $h_2 = h \exp(-\beta\delta)$ for aggregation. Specifically, $h_1$ and $h_2$ describe the rate of the last hop leading to nucleation and aggregation, respectively. Since $\delta \geq 0$ and $\delta^* \geq 0$, $h_1$ and $h_2$ are generally reduced below $h_0 = h$ corresponding to inhibited nucleation and aggregation, and we set $r_j \equiv h_j/h \leq 1$, for $j = 1$ or 2. It is sometimes useful to introduce $L_{\delta^*} = \exp(\beta\delta^*) = 1/r_1$ and $L_\delta = \exp(\beta\delta) = 1/r_2$ denote "attachment lengths" (in units of the surface lattice constant) for nucleation and aggregation, respectively, larger attachment lengths implying more difficult attachment processes [29]. Note that $r$ (= $r_1$ = $r_2$) = 1 corresponds to the classic diffusion-mediated PIM considered in Sec.2 and Sec.3. Here and below, we always use $r$ for $r_1 = r_2$. A key feature of these models, discussed further below, will be that increasing attachment barriers (or reducing $r_j$) makes the adatom density more spatially uniform relative to the diffusion-limited regime $r = 1$.

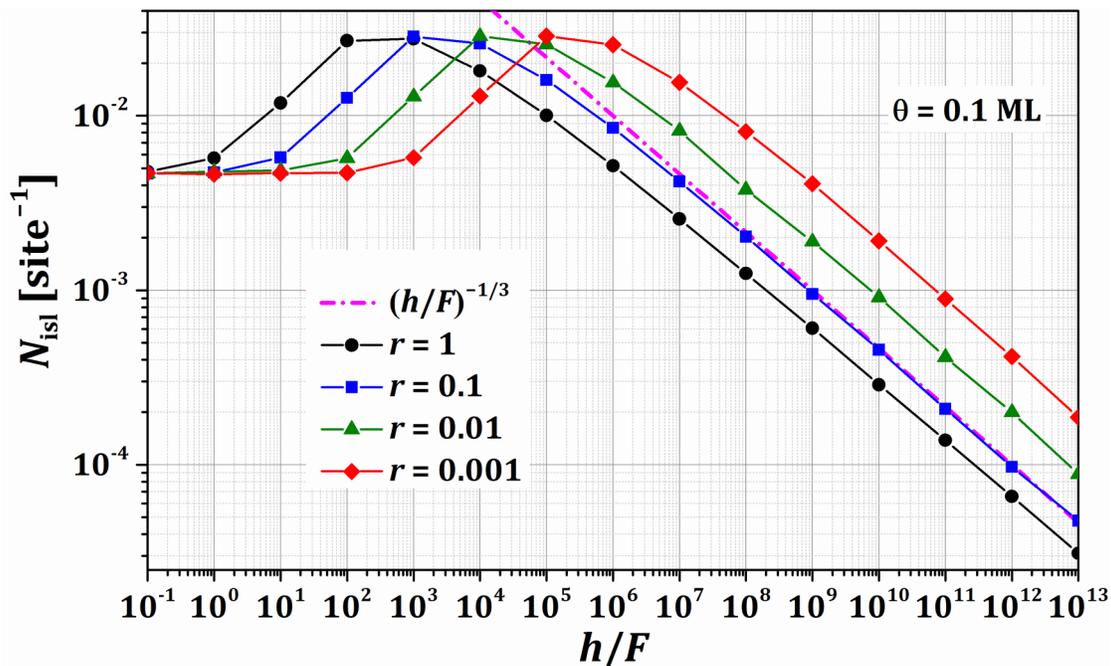

**Fig.7**. KMC results for $N_{isl}$ versus h/F for the PIM with various additional barrier $\delta = \delta^*$ so that $r = \exp(-\beta\delta) \leq 1$.

One benchmark case of the model sets $\delta = \delta^*$ so that $h_1 = h_2$, and $r = \exp(-\beta\delta) \leq 1$. KMC simulation results for $i = 1$ shown in Fig.7 indicate that the scaling behavior $N_{isl} \sim (h/F)^{-1/3}$ familiar in the classic PIM for $r = 1$ is actually preserved even for $r << 1$. This behavior is explained below in Sec.4B. The other key feature of the simulation results is that $N_{isl}$ increases with increasing barrier $\delta$ (or with decreasing $r$) at fixed $h/F$. Note that the increase in $N_{isl}$ with increasing $\delta$ or decreasing $r$ helps avoid corruption due to finite-size effects in simulation for large $h/F$. Just increasing the barrier for nucleation would reduce $N_{isl}$, but increasing the barrier for aggregation will have the opposite effect boosting the adatom density and thus the nucleation rate. For $\delta = \delta^*$, the latter effect dominates. This feature has been observed in previous experimental and simulation studies mentioned in Sec.6. A quantitative analysis of behavior for fixed $h/F$ and increasing δ based on the results described below in Sec.4B suggests that $N_{isl}$ ~ exp[βδ/3] = $r^{-1/3}$ for δ = δ*. Our simulation data is consistent with this behavior.

In addition, we explore evolution of the ISD for increasing attachment barrier, i.e., for decreasing $r$, where we simultaneously increase $h/F$ to maintain a roughly constant $N_{isl}$. Results shown in Fig.8 suggest that the shape evolves towards the singular MF form (13) for $i$ = 1 described in Sec.2A. This behavior is naturally understood from the feature that the more uniform adatom density leads to adatom capture which is more independent of island size $s$. This in turn implies that appropriate rate equation analysis will naturally produce MF-type behavior. See Sec.4B for further discussion.

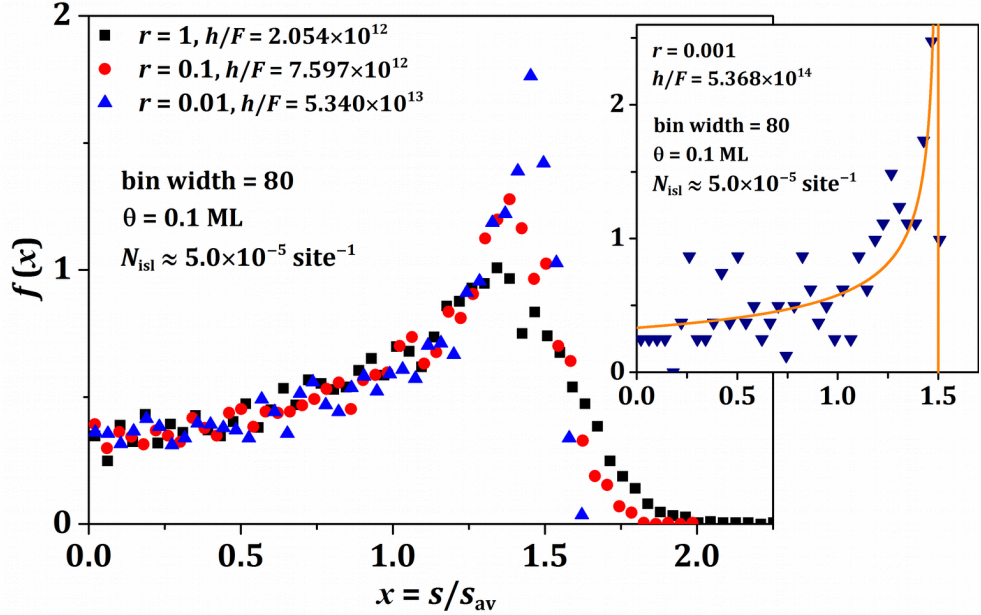

**Fig.8**. KMC results for the ISD for the PIM with additional barrier δ = δ* so that $r$ = exp(-βδ) ≤ 1 adjusting h/F to maintain a constant $N_{isl}$ ≈ 5×$10^{-5}$. The inset also superimposes the proposed MF scaling function upon noisy simulation data for the smallest $r$ = 0.001.

Finally, we discuss a second benchmark case of the model setting δ* = 0 so that $h_1$ = $h$, but δ ≥ 0 so that $h_2$ ≤ $h$ and $r_2$ = exp(-βδ) ≤ 1, i.e., we inhibit aggregation but not nucleation. In this case, it is obvious that $N_{isl}$ will increase with increasing δ, and the prediction of the formulation in Sec.4B is that $N_{isl}$ ~ exp[2βδ/3] = $(r_2)^{-2/3}$ for δ* = 0. We have checked that our simulation results are consistent with this behavior. For this type of model, there is the general and reasonable perception that inhibiting aggregation by inclusion of the additional barrier will lead to more persistent nucleation relative to the standard model with $r_2$ = 1. This, in turn, suggests that the variation of $N_{isl}$ with coverage, θ, should become closer to linear in contrast to the strongly sublinear behavior quantified by (3c) for $r_2$ = 1. Clearly, an infinite barrier, δ = ∞, allows formation only of dimers, so one has that $N_{isl}$ ≈ θ/2. Indeed, KMC simulation results for $i$ = 1 shown in Fig.9 reveal that increasing $\delta$ (or decreasing $r_2$) does ultimately produce linear behavior, but this transition only occurs "very slowly". For expected typical physical magnitudes of barriers where $\beta\delta \leq 4$ (e.g., $\delta \leq 0.1$ eV at 300 K), it is clear that strongly non-linear variation of $N_{isl}$ with $\theta$ should still be observed.

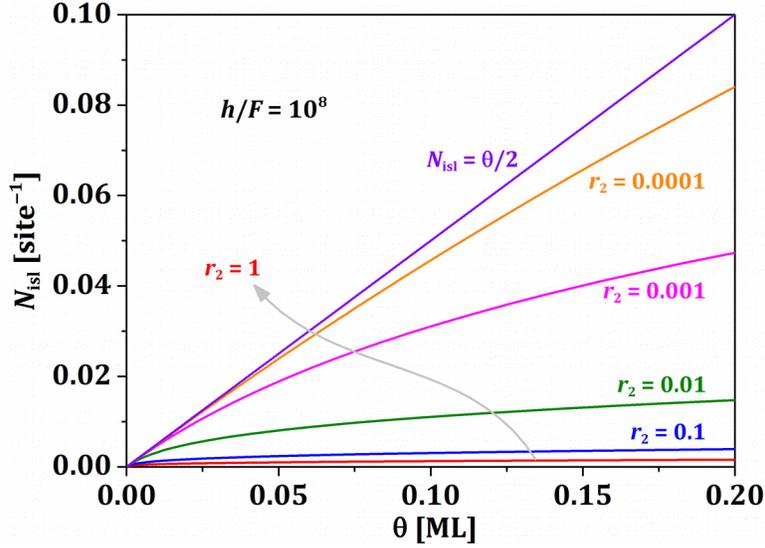

**Fig.9**. Coverage-dependence of $N_{isl}$ for inhibited aggregation ($r_1 = 1$ and $r_2 \leq 1$).

### 4B. PIM WITH EXTRA BARRIERS MIMICING 2D AND 3D ISLANDS

In this subsection, we focus on the regime of large additional barriers for nucleation and growth, so that these processes are attachment- rather than diffusion-limited. This, in turn, means that the adatom density, $N_1$, will be relatively uniform across the surface. We will exploit this feature to develop an analytic theory for island nucleation and growth which is appropriate for real 2D ($d = 2$) and 3D ($d = 3$) island systems, as well as for standard point islands. To this end, we note that if $P_s$ denotes the perimeter length for an island of size $s$, then the aggregation rate for that island for quasi-uniform adatom density, $N_1$, is given by

$$k_{agg}(s) = h \exp(-\beta\delta) P_s N_1. \qquad (20)$$

It is clear that $P_s \sim 1$ for point islands, and that $P_s \sim s^{1/d}$ for 2D and 3D islands. Note that one can also use the latter form for point islands by assigning $d = \infty$.

The form (20) suggests a simple refinement of the prescription of aggregation with large extra barriers within the PIM framework in order to mimic behavior in real 2D and 3D island systems. This refined PIM retains a hop rate $h_0 = h$ for terrace diffusion, and $h_1 = h \exp(-\beta\delta^*)$ for nucleation, but now assign $h_2 = h_2(s) = h \exp(-\beta\delta)s^{1/d}$ for aggregation. This assignment reflects the enhancement of aggregation rate with increasing island size and thus perimeter length. Certainly for $\exp(-\beta\delta)s^{1/d} < 1$, this formulation with $d = 2$ or $d = 3$ will incorporate the enhanced propensity for adatom capture at larger islands [30]. As an aside, we note that this refined PIM is even better suited for analysis of 3D islands than of 2D islands since the "footprint" of the former is smaller, or more point-like, at a fixed coverage. Thus, the coalescence regime occurs at higher coverage for 3D islands, and the refined PIM accurately describes behavior for a broader range of coverage.

Before presenting simulation results for these refined PIM, we develop an appropriate analytic treatment for their behavior. To this end, we consider the average aggregation rate <u>per island</u>, $k_{agg} = \exp(-\beta\delta) h P_{av} N_1$, where the average perimeter length satisfies

$$P_{av} \sim (\theta/N_{isl})^{1/d} \sim \theta^{1/d} (L_{isl})^{2/d} \text{ for } d = 2, 3, \text{ or } \infty. \tag{21}$$

In the steady-state regime, $N_1$ is determined by a balance between the rate of gain <u>per CZ</u> due to deposition, $k_{dep} \sim F(L_{isl})^2$, and the rate of loss, $k_{agg}$, due to aggregation with the island in that CZ. This implies that

$$N_1 \approx \exp(+\beta\delta)(h/F)^{-1}(L_{isl})^2/P_{av} \sim \theta^{-1/d} \exp(+\beta\delta)(h/F)^{-1} (N_{isl})^{-(d-1)/d}. \tag{22}$$

With this result, we can determine the nucleation rate,

$$K_{nuc} = \sigma_i \exp(-\beta\delta^*-\beta E_i) h (N_1)^{i+1}. \tag{23}$$

Then, integration of $d/dt\, N_{isl} = K_{nuc}$ yields the expression for the island density

$$N_{isl} \sim [\sigma_i \exp[-\beta E_i - \beta\delta^*+(i+1)\beta\delta] \int d\theta\, \theta^{-(i+1)/d} ]^{1/[(i+2)-(i+1)/d]} (h/F)^{-i/[(i+2)-(i+1)/d]}. \tag{24}$$

For the integral determining the θ-dependence, one should just integrate over the steady state regime for $\theta \geq \theta^*$ for some suitable initial conditions.

Focusing on the scaling with $h/F$, we obtain

$$N_{isl} \sim (h/F)^{-\chi} \text{ with } \chi = i/(i+2) \text{ for } d = \infty, \chi = 3i/(2i+5) \text{ for } d = 3, \text{ and } \chi = 2i/(i+3) \text{ for } d = 2. \tag{25}$$

The results for the standard PIM ($d = \infty$) and for the refined PIM mimicking 3D islands ($d = 3$) are new, and those for the refined PIM mimicking 2D islands ($d = 2$) are found in Ref.[27]. The standard PIM reveals the same scaling with h/F and with θ as for diffusion-mediated growth with $\delta = \delta^* = 0$. The scaling $N_{isl} \sim (h/F)^{-1/3}$ for PIM with $i = 1$ is clearly illustrated in Sec.4A.

The study in Ref.[27] in fact performed a more comprehensive analysis for 2D islands assessing the <u>crossover</u> from diffusion-limited to attachment-limited growth with increasing magnitude of the additional barriers. The basic idea, which extends to other island structures, is that crossover should occur from the diffusion-limited to the attachment-limited regime as the relevant attachment length, $L_{attach}$, increases beyond the mean island separation, $L_{isl}$. Another perspective is that for fixed high barriers, and thus fixed $r_i \ll 1$, one expects a transition from attachment-limited to diffusion-limited behavior with increasing $h/F$ as $L_{isl}$ grows to exceed $L_{attach}$. For the regime of modified scaling where $L_{isl} \ll L_{attach}$, we do not expect significant finite-size effects in an $L \times L$ system for $L < L_{attach}$, provided that L is significantly larger than $L_{isl}$.

Next, we comment briefly in the expected form of the ISD for large attachment barriers. For standard point islands ($d = \infty$) where $P_s \sim 1$, as noted above, uniformity of the adatom density, $N_1$, implies that islands will capture adatoms at the same rate independent of size. Thus, the rate equations (9) for the densities, $N_s$, of islands of size $s$ will apply with $K_{agg}(s) = \sigma_s\, hN_sN_1$ and $\sigma_s = \sigma \exp[-\beta\delta]$ independent of $s$. This in turn indicates that the MF results (13) described in Sec.2B will be applicable. For 2D and 3D islands, one instead has that $K_{agg}(s) \propto h \exp[-\beta\delta]\, P_s\, N_sN_1$, where $P_s \sim s^{1/d}$. This feature modifies the form of the ISD as explored in previous analyses [31,32].

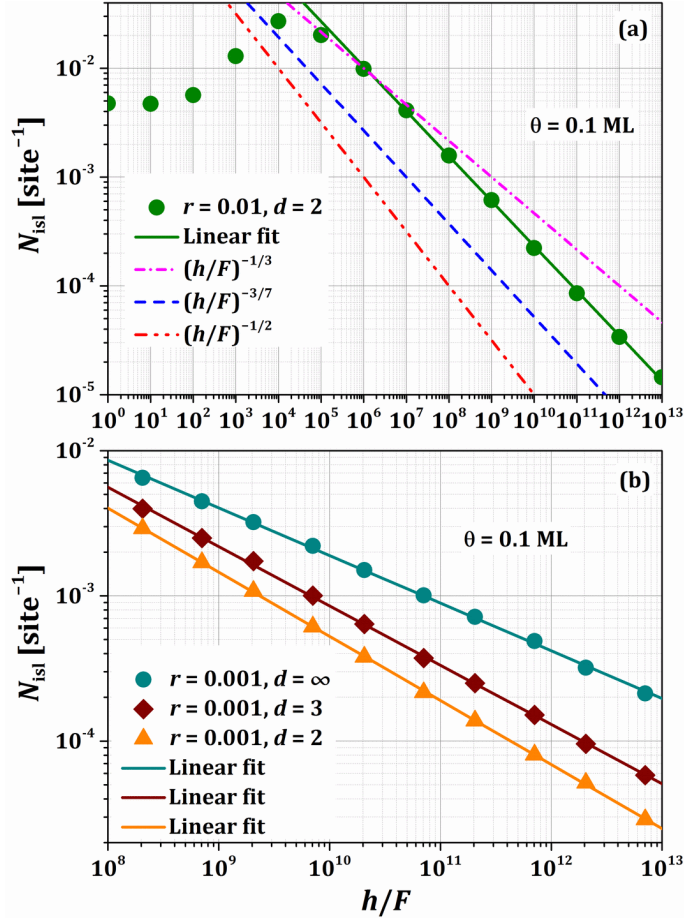

**Fig.10**. Analysis of scaling, $N_{isl} \sim (h/F)^{-\chi}$, for the attachment-limited regime with large barriers in PIM mimicking 2D and 3D islands. (a) Comprehensive characterization of $N_{isl}$ for 2D islands for $r = 0.01$; $\chi \approx 0.42$ for $h/F$ from $10^7$ to $10^{12}$. (b) Scaling of $N_{isl}$ for $r = 0.001$ for point ($d = \infty$), 3D ($d = 3$), and 2D ($d = 2$) islands. We find that $\chi \approx 0.33$ (point), $\chi \approx 0.41$ [vs. 3/7 from (35)] (3D), $\chi \approx 0.44$ [vs. 1/2 from (35)] (2D) for the range shown.

Finally, we describe selected KMC simulation results for our modified PIM mimicking 2D and 3D islands for $i = 1$ and with large $\delta = \delta^*$, so that $r\ (= r_1 = r_2) \ll 1$. See Fig.10. The goal is to probe modified scaling with h/F relative to behavior for point islands. For $r = 0.01$, the attachment length, $L_\delta \approx 100$, is well above the mean island separation, $L_{isl}$, for h/F up to about $10^{12}$ from Fig.7. Thus, modified scaling should be found for such $h/F$. From Fig.10a, we estimate that $\chi \approx 0.42$ for $h/F$ from $10^7$ to $10^{12}$ for 2D islands [significantly below $\chi \approx -1/2$ from (25)]. Similar analysis for 3D islands yields $\chi \approx 0.39$ [versus $3/7 = 0.428$ from (25)]. Plausibly, scaling exponents will be closer to analytic estimates for smaller $r$ = 0.001 corresponding to $L_\delta$ = 1000. This $L_\delta$ is far above the average island separation for all $h/F$ accessed by simulation. From the data in Fig.10b, we find that $\chi \approx$ 0.44 for 2D islands, and $\chi \approx$ 0.41 for 3D islands (versus $\chi \approx$ 0.33 for point islands).

For the above small r, it is challenging to precisely quantify crossover from modified scaling for 2D or 3D islands for smaller $h/F$ to scaling with $\chi \approx$ 1/3 for large $h/F$ (just because it is difficult to obtain precise data for such large $h/F$). One possibility is to select a larger $r$ = 0.04, say, so that crossover should occur for smaller $h/F$ likely around $10^8$ to $10^9$. In this case for 2D islands with $r$ = 0.04, we do find a transition from higher values of $\chi \approx$ 0.39 for $h/F$ from $10^6$ to $10^9$, say, to lower values of $\chi \approx$ 0.34 for $h/F$ from $10^{11}$ to $10^{14}$.

## 5. PIM FOR MODIFIED NUCLEATION AND GROWTH PROCESSES

In this section, we highlight the versatility of PIM for treating a diverse variety of mechanisms as well as other possible features of nucleation and growth systems.

(i) <u>Anisotropic diffusion</u>. It is straightforward to modify the standard model to include anisotropic or even 1D diffusion [11]. In the latter extreme 1D case, one must specify if nucleation and aggregation only occur with adatoms and clusters in the same row (version A) or also in the adjacent rows (version B) along which diffusion occurs. For strong anisotropy, the basic scaling behavior of the island density [11,33],

$$N_{isl} \sim \exp[2\beta E_i/(i+2)](h/F)^{-\chi} \qquad (26)$$

with $\chi$ = 2/(2$i$+1), for critical size $i$, is distinct from the isotropic case for both versions A and B (which exhibit similar behavior for large $h/F$). This behavior is obtained from modified rate equations accounting for the distinct nature of diffusion in 1D as indicated in Appendix A. Crossover from isotropic to strongly anisotropic behavior in PIM can be analyzed and quantified [34].

(ii) <u>Diffusion of small stable clusters</u> also impacts scaling of $N_{isl}$ in nucleation and growth if diffusivity is comparable to that on single adatoms. Here one introduces a new parameter, $i*$, wherein clusters of size i* and smaller are mobile. Thus, the case $i$ = 1 and $i*$ = 2 corresponds to irreversible island formation with mobile dimers. For $i$ = 1 and general $i*$ where all mobile clusters have hop rates comparable to $h$, a rate equation analysis reveal that $N_{isl} \sim (h/F)^{-\chi}$ with $\chi$ = $i*/(2i*+1)$ [35,36]. The basic trend in behavior is that increasing $i*$ allows greater incorporation of small stable mobile islands with larger stable immobile islands, and thus $N_{isl}$ decreases. These models can be naturally implemented within a PIM framework specifying desired hop rates for clusters with sizes $s \leq i*$. We have implement these models for $i$ = 1 and $i*$ = 2 and 3 (in addition to the standard model for irreversible island formation with $i$ = $i*$ = 1). Results shown in Fig.11, for the case where all mobile clusters have the same hop rate, $h$, as that of adatoms, are consistent with the above scaling law, but also illustrate the extent of decrease in $N_{isl}$ produced by small cluster mobility.

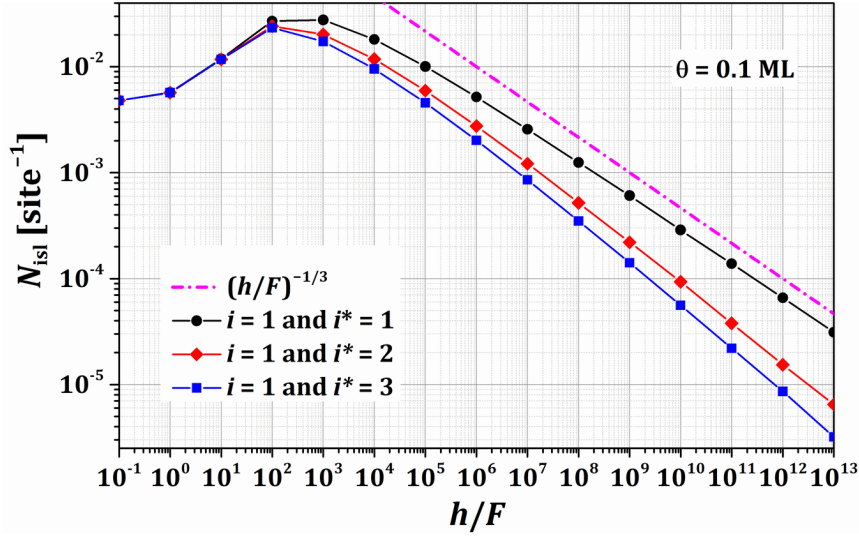

**Fig.11**. $N_{isl}$ (including all islands with $s \geq 2$) versus $h/F$ for PIM with $i = 1$ and $i^* = 1 - 3$ at 0.1 ML (with equal hop rate h for all mobile adspecies). Exponents from this data for h/F from $10^9$ to $10^{13}$ are $\chi = 0.32$, 0.38, and 0.41 for $i^* = 1$, 2, and 3, respectively.

(iii) <u>Long-range interactions</u>. In some systems, interactions between adatoms are not just dominated by very short-ranged attractions responsible for NC formation [28,37] (as assumed above). Also, any additional barriers inhibiting attachment may not just impact the last hop before nucleation or aggregation, but also hopping rates for further separated adatoms. The simplest scenario for implementation of long-range interactions is to modify PIM model type A (where implicitly there are zero range attractions since nucleation and aggregation occur when an adatom hops onto the same site as another adatom or island). In this case, we specify an additional longer-range interaction, $\phi_1(j,k)$ between adatoms at adsorption sites with separation $\underline{r}_s = (j,k)$ in units of surface lattice constant for $r_s = |\underline{r}_s| > 0$, e.g., repulsive $\phi_1(r_s) = \varepsilon/[1+(r_s/r_c)^n]$ for $\varepsilon > 0$, where $r_c$ measures the interaction range [28]. Distinct interactions $\phi_s(j,k)$ are prescribed between an adatom and island of size $s$.

The key ingredient for models of nucleation and growth is to specify barriers for all hops. For a diffusing adatom, one can determine the total interaction $\Phi_{tot}$ with other islands and adatoms by summing the above pair interactions, $\phi$, for both the initial (init) and final (fnl) states before and after hopping. In practice for large h/F and small $i$, one expects the dominant contribution to come from one $\phi$ associated with a single nearby island or adatom. One can write the diffusion barrier as $E_{act} = E_d + \Phi_{tot}(TS) - \Phi_{tot}(init)$. Here, $\Phi_{tot}(TS)$ is the total interaction at the transition state for hopping which is in principle <u>not</u> determined by the above $\phi$. For slowly varying $\phi$, one can reasonably utilize a symmetric Brønsted–Evans–Polanyi type approximation, $\Phi_{tot}(TS) = \frac{1}{2} [\Phi_{tot}(init) + \Phi_{tot}(fnl)]$ [38,39].

(iv) <u>Growth inhibition</u>. For heteroepitaxial growth, a lattice mismatch between the substrate and the crystalline form of the deposited material leads to strain buildup, and thus a significant energetic penalty for larger islands inhibiting their growth [40]. A simple strategy to model this behavior is to reduce the rate of the last adatom hop leading to aggregation in the PIM for larger islands by a factor of $f(s) \leq 1$. Specifically, one chooses a threshold size $s^*$ for growth inhibition, and sets $f(s) \approx 1$ for $s \ll s^*$ and $f(s) \approx 0$ for $s \gg s^*$. A simple default choice is $f(s) = \tanh[(s-s^*)/w]$. For this model, one can also include an additional barrier, $\delta$, for nucleation and aggregation so those rates are reduced by an additional factor $r = \exp(-\beta\delta)$. In Fig.12, we show simulation results for a modified PIM with $i = 1$, $h/F = 3 \times 10^{13}$, $s^* = 1000$ for various w for both $r = 1$ and $r = 0.01$. Compared to models with no growth inhibition factor $f(s)$, there is the "ready development" of a linear increase of $N_{isl}$ with $\theta$, noting that $s_{av} = \theta/N_{isl}$ quickly exceeds $s^*$ for the above choice of model parameters.

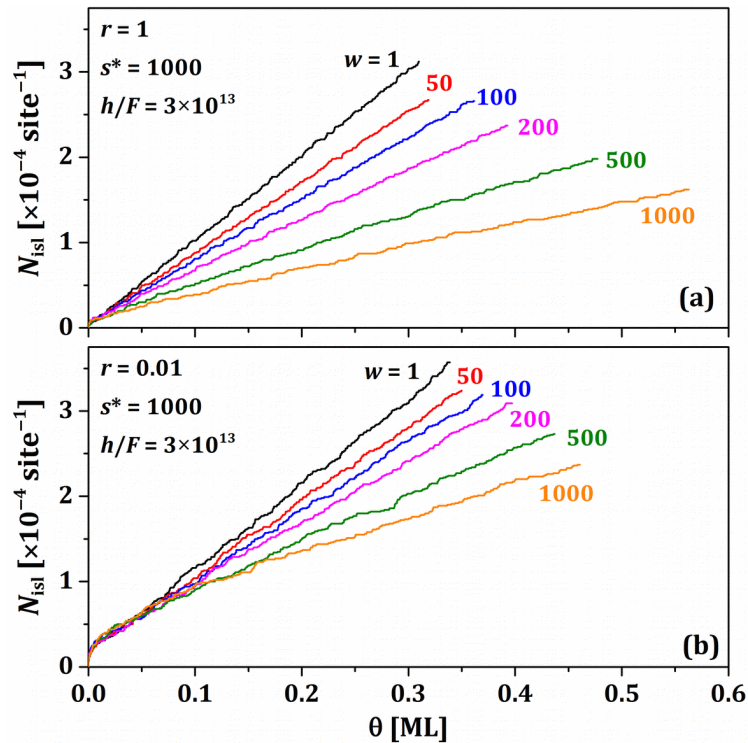

Fig.12. $N_{isl}$ versus $\theta$ for irreversible island formation with growth inhibition.

(v) <u>Nucleation mediated by surface exchange ($i = 0$)</u>. A distinct nucleation mechanism has been suggested for some heteroepitaxial metal-on-metal systems [41,42]. Here, deposited adatoms can exchange with atoms in the top surface layer thereafter remaining essentially immobile and embedded in that layer. Furthermore, these embedded adatoms provide nucleation centers for subsequently deposited adatoms. This nucleation scenario has been labeled as $i = 0$ since the single embedded adatom forms a stable nucleus for subsequent island growth. A PIM for this process has been implemented previously. In addition, rate equations were also developed to describe a potentially unusual variation of island density with temperature [42].

(vi) <u>Sequential (or simultaneous) codeposition</u>. For sequential codeposition of species with strongly differing surface mobilities, the nucleation and growth process depends strongly on the order of deposition [43,44]. If the less mobile species is deposited first, a higher density of NCs of that species is formed which provide nucleation centers during the second stage of deposition. This results almost exclusively in two-component (possibly core-shell) NCs after the second deposition. If the more mobile species is deposited first, then a lower density of pure NCs of that species is formed which allows nucleation of new pure NCs of the less mobile species in the second stage of deposition. In addition, NCs formed in the first stage of deposition are mainly converted to two-component NCs. PIM have been implemented for these processes, also incorporating additional features described in (ix) below [44]. PIM can also readily treat the case of simultaneous codeposition.

(vii) <u>Heterogeneous nucleation at preexisting defects</u>. Naturally, PIM can be adapted to describe heterogeneous (rather than homogeneous) nucleation on surfaces where preexisting immobile defects provide nucleation centers. However, behavior for these processes is relatively trivial compared to homogeneous nucleation. It is natural to tessellate the surfaces into CZs surrounding each defect, where in the simplest case tessellation cells can be approximated by Voronoi cells. Then, it is clear that island growth rates are controlled by the corresponding CZ areas, so that the ISD should trivially match the CZ area distribution [45] (in contrast to the more complicated and subtle behavior for homogeneous nucleation) [9].

(viii) <u>Heterogeneous nucleation at deposition-induced defects</u>. For deposition using an e-beam evaporator, some typically small fraction of the depositing species can be ions rather than neutral atoms. These ions can damage more fragile substrates, thereby creating nucleation centers during the deposition process [46]. Such deposition centers will be created and at constant rate and formed at randomly locations on the surface. Thus, the nucleation process is effectively described by a Johnson-Mehl-Avrami-Kolmogorov model [10] potentially modified to account for a variable rate for radial island growth in this model [46]. PIM have been recently modified to treat this type of nucleation process [46].

(ix) <u>Directed-assembly</u>. For either homogeneous nucleation, or heterogeneous nucleation at preexisting defects, there is effectively no control of nucleation locations. This shortcoming is addressed in directed assembly, where nucleation locations are guided by a templated substrate [47]. Consider a substrate which is periodically templated or modulated in the sense that some aspect of the adsorption energetics varies periodically on some coarse length scale, $L_c$ (i.e., $L_c$ is significantly larger than the surface lattice constant). Here, there is the potential to induce the formation of periodic arrays of NCs. PIM have been implemented for such assembly which can be thermodynamically-directed by modulating the adsorption energy, or kinetically-directed by modulating the diffusion barrier (or both) [48].

## 6. DISCUSSION AND CONCLUSIONS

The above presentation is intended to highlight into the value and utility of point island models (PIM) for elucidation of: fundamental issues in homogeneous diffusion-limited nucleation and growth (Sec.2 and 3); distinct behavior in the regime of attachment-limited nucleation and growth (Sec.4); and the effect of a diverse variety of modified mechanisms and features of nucleation and growth processes during deposition on surfaces (Sec.5).

We have emphasized model development and analysis rather than applications. However, PIM can be applied directly to elucidate and interpret behavior in a host of specific systems, some of which are indicated above. An expanded list of examples includes: (a) homogenous nucleation and growth of metal NCs during deposition on metal, semiconductor, oxide substrates [8,9], with particular utility for Volmer-Weber growth of 3D islands on weakly binding substrates; (b) nucleation inhibited by attachment barriers as observed in metal (111) homoepitaxial systems which exhibit enhanced long-range adatom interactions with a "repulsive ring" due to surface states [38,49]; inhibited attachment was also recently suggested for Fe deposition on graphene [50]; (c) nucleation and growth with strongly anisotropic diffusion as observed for homoepitaxy on dimer-row reconstructed Si(100) surfaces [51]; (d) significant effects on nucleation of small mobile clusters as anticipated for homoepitaxy on metal (100) and (111) surfaces [52]; (e) growth inhibition in strained-layer heteroepitaxy with large mismatch [53]; (f) codeposition to form bimetallic NCs, e.g., of Pt and Au on $TiO_2$(110) [43], and Pt and Ru on graphene [44]; (g) exchange-mediated nucleation in Fe on Cu(100) [41], and Ni on Ag(111) [42] systems; (h) nucleation of metal NCs on graphite is often facilitated by sputtering to create surface damage and heterogeneous nucleation centers, but even a small fraction of Cu ions from an e-beam evaporator can create sufficient damage that heterogeneous nucleation dominates for Cu deposition on HOPG [46]; (i) deposition of metals on metal-supported graphene, with periodically rumpled morié structure due to lattice mismatch, often results in directed-assembly of 3D NCs [54], and the PIM framework is ideally suited to modeling of this complex process [48].

Finally, all of the above discussion of PIM analysis pertains to nucleation and growth of NCs during deposition. We should mention that the PIM approach can be extended to treat post-deposition coarsening of NC arrays [55]. Coarsening can occur either by cluster diffusion and coalescence (where one would incorporate appropriate size-dependent diffusion coefficients for NCs) or via Ostwald ripening (where one would incorporate an appropriate size-dependent detachment rate of adatoms from NCs).

## ACKNOWLEDGEMENTS

YH and JWE were supported for this work by NSF grants CHE-1111500 and CHE-1507223. The work was performed at Ames Laboratory which is operated for the USDOE by Iowa State University under Contract No. DE-AC02-07CH11358. Computations utilized USDOE NERSC, OLCF, and NSF-supported XSEDE resources. EG was granted access for this work to the HPC resources of GENCI (Grand Equipement National de Calcul Intensif) under the allocation 96339. TJO acknowledges the support from CNPq and FAPEMIG (Brazilian agencies).

## APPENDIX A: FORMULATIONS BASED ON RANDOM WALK THEORY

Rather than the standard MF expressions for diffusion-mediated nucleation and aggregation rates in Sec.2, an alternative more fundamental and flexible formulation is based on the lifetime, $\tau$, of deposited adatoms [9,11,33,34]. If $N_t$ is the density of "traps" for a diffusing adatom, so $N_t = N_{isl} + N_i \approx N_{isl}$ in the steady-state, where again $N_{isl}$ is the density of stable islands and $N_i$ is the density of critical clusters. Then, the mean number of hops, $m_t$, of a deposited atom before trapping through nucleation or aggregation satisfies $m_t = h\tau \sim (N_t)^{-1}\ln[(N_t)^{-1}]$ for isotropic diffusion in 2D, and $m_t = h\tau \sim (N_t)^{-2}$ for 1D diffusion. Then, the rate for aggregation with NCs of size s are given by $K_{agg}(s) \approx N_1 q_s/\tau$ where $q_s = N_s/N_t$ is the probability to aggregate with an island of size s. Similarly, $K_{nuc} = K_{agg}(i)$ and $K_{agg} \approx N_1/\tau$. Matching these results with MF expressions, we conclude that $\sigma_s \sim \sigma_i \sim \sigma_{av} \sim 1/\ln[(N_t)^{-1}]$ for isotropic 2D diffusion, but $\sigma_s \sim \sigma_i \sim \sigma_{av} \sim N_t$ for 1D diffusion. The former result yields just log-corrections to the previously derived results for the 2D case, but fundamentally different scaling behavior for 1D diffusion.

## APPEDIX B: FURTHER ANALYSIS OF FINITE-SIZE EFFECTS

For the case of diffusion-limited homogeneous nucleation and growth (without additional barriers for nucleation and aggregation), we further quantify finite-size effects on the ISDs by analyzing the behavior of the first four cumulants of ISDs versus $L/L_{isl}$. Results in Fig.13 show clear scaling with $L/L_{isl}$, but slower convergence to limiting behavior at $L/L_{isl} \approx 20\text{-}30$ than observed for $N_{isl}$. This is reasonable since the ISD is expected to be more sensitive to finite-size effects than the mean density, $N_{isl}$.

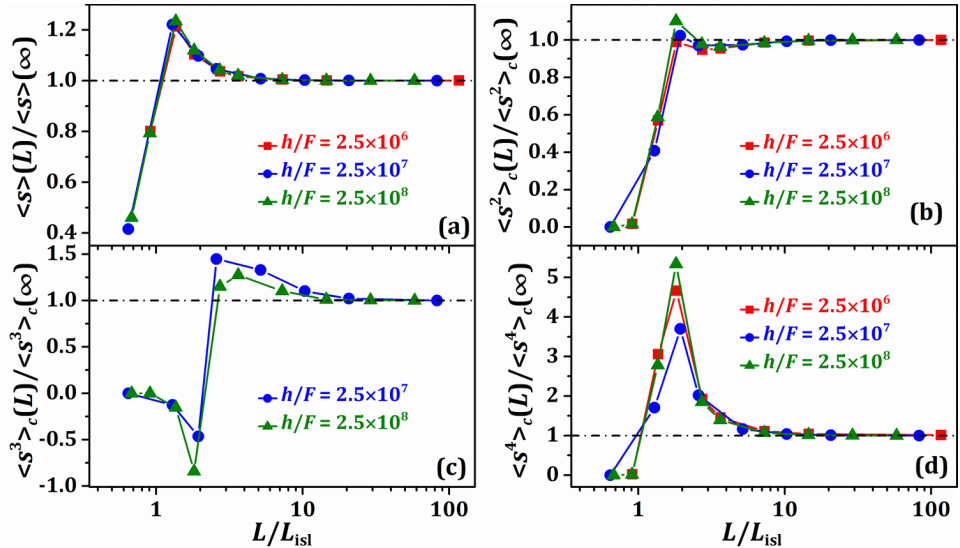

**Fig.13.** Variation with $L/L_{isl}$ of the rescaled cumulants $<s^n>_c(L)/<s^n>_c(\infty)$ of ISDs for (a) $n = 1$, (b) $n = 2$, (c) $n = 3$, and (d) $n = 4$. Data for version (B) of the PIM with $i = 1$.

Next, we expand the analysis in Sec.3B of the few-island regime where $L/L_{isl}$ is of order unity. We have already assessed the probability for nucleation of the <u>first and second island</u> as a function of coverage. To assess the existence of the <u>third island</u>, one should convolute the probability density $dP_2/d\theta' = F^{-1} K_{Tnuc1} \exp[-\theta' K_{Tnuc1}/F]$ for the second island to be formed at coverage $\theta' < \theta$ with the probability $1 - \exp[-\delta\theta K_{Tnuc2}/F]$ that the third island has formed after an additional coverage $\delta\theta = \theta - \theta'$. Here, $K_{Tnuc2}$ is the time-independent total nucleation rate for the third island in the presence of two existing islands. Actually, $K_{Tnuc2}$ depends on the position of the second island relative to the first. Thus, one should really account these island-position-dependent rates, but we ignore this complication. Evaluating the convolution integral, we obtain

$P_3(\theta) \approx \{1 - \exp(-\theta K_{Tnuc1}/F)\}$

$- K_{Tnuc1}(K_{Tnuc1} - K_{Tnuc2})^{-1} \exp(-\theta K_{Tnuc2}/F) \{1 - \exp[-\theta(K_{Tnuc1} + K_{Tnuc2})/F]\}$.  (B1)

Note that $P_3$ initially increases quadratically with $\theta$ (in contrast to $P_2$ which initially increases linearly). Below, we will exploit the feature that $K_{Tnuc2}$ has the same scaling as $K_{tnuc1}$, but is just reduced by a factor of $p_2 < 1$ since the steady-state $N_1$ is reduced on the more crowded surface. Setting $K_{Tnuc2} = p_2 K_{Tnuc1}$ with $p_2 < 1$. Then

$N_{isl}(L)/N_{isl}(\infty) = g(L/L_{isl})$, where

$g(u) \approx u^{-2} \{1 + 2[1 - \exp(-b\theta u^{2(i+2)})] - (1-r_2)^{-1}\exp[-bp_2 \theta u^{2(i+2)}](1 - \exp[-b(1-p_2)\theta u^{2(i+2)}])\}$. (B2)

The data in Fig. 6b with $\theta = 0.1$ ML around $L/L_{isl} \approx 1$ and is fit choosing $b = 1.1$, $p_2 = 0.75$.

As a final aside, we note that analysis of finite size effects is particularly relevant for deposition on vicinal surfaces, where a transition from island formation on terraces to step flow occurs for terrace widths $L_{terr} \sim L_{isl}$ [56]. We expect that similar crossover behavior occurs to that described above.